\documentclass[10pt,aps,pra,twocolumn,superscriptaddress,floatfix,nofootinbib]{revtex4-2}
\makeatletter
\def\@bibdataout@aps{
 \immediate\write\@bibdataout{
  @CONTROL{
   apsrev41Control,author="08",editor="1",pages="0",title="0",year="1",eprint="1"
  }
 }
 \if@filesw
  \immediate\write\@auxout{\string\citation{apsrev41Control}}
 \fi
}
\makeatother

\usepackage{lipsum}
\usepackage{amsmath}
\usepackage{amsfonts, amssymb, bbm, braket, accents}
\usepackage{graphicx}
\usepackage{wrapfig}
\usepackage[usenames,dvipsnames]{color}
\usepackage[makeroom]{cancel}
\usepackage{soul}
\usepackage{tikz}
\usepackage{float}
\makeatletter
\let\newfloat\newfloat@ltx
\makeatother
\usepackage{algorithm}
\usepackage{algpseudocode}
\usetikzlibrary{shapes.arrows} 
\usetikzlibrary{shapes.geometric}
\usetikzlibrary{calc} 
\usepackage[makeroom]{cancel}
\usepackage{verbatim}
\usepackage{pifont}
\usepackage[mathscr]{euscript}

\DeclareMathOperator*{\argmin}{argmin}

\usepackage{booktabs}
\usepackage{xcolor}
\usepackage{siunitx}
\usepackage{colortbl}
\usepackage{algorithm}
\usepackage{algpseudocode}

\definecolor{cellmin}{rgb}{1,1,1}
\definecolor{cellmax}{rgb}{0.25,1,0.25}

\newcommand{\probP}{\text{I\kern-0.15em P}}
\newcommand{\probE}{\text{I\kern-0.15em E}}

\usepackage[breaklinks=true]{hyperref}
\hypersetup{
  colorlinks   = true,
  urlcolor     = blue,
  linkcolor    = blue,
  citecolor   = red
}
\usepackage{amsthm}

\usepackage[capitalise]{cleveref}
\crefformat{equation}{Eq.~(#2#1#3)}
\crefformat{section}{Sec.~#2#1#3}
\Crefformat{equation}{Equation~(#2#1#3)}
\crefformat{figure}{Fig.~#2#1#3}
\crefrangeformat{equation}{Eqs.~#3(#1)#4--#5(#2)#6}
\Crefformat{section}{Section~#2#1#3}

\begin{document}
\title{On the Interpretability of Quantum Neural Networks}

\author{Lirandë Pira}
\email{lirande.pira@student.uts.edu.au}
\affiliation{University of Technology Sydney,
		Centre for Quantum Software and Information,
		Ultimo NSW 2007, Australia}

\author{Chris Ferrie}
\affiliation{University of Technology Sydney,
		Centre for Quantum Software and Information,
		Ultimo NSW 2007, Australia}

\date{\today}

\begin{abstract}
Interpretability of artificial intelligence (AI) methods, particularly deep neural networks, is of great interest. This heightened focus stems from the widespread use of AI-backed systems. These systems, often relying on intricate neural architectures, can exhibit behavior that is challenging to explain and comprehend. The interpretability of such models is a crucial component of building trusted systems. Many methods exist to approach this problem, but they do not apply straightforwardly to the quantum setting. Here, we explore the interpretability of quantum neural networks using local model-agnostic interpretability measures commonly utilized for classical neural networks. Following this analysis, we generalize a classical technique called LIME, introducing Q-LIME, which produces explanations of quantum neural networks. A feature of our explanations is the delineation of the region in which data samples have been given a random label, likely subjects of inherently random quantum measurements. We view this as a step toward understanding how to build responsible and accountable quantum AI models.
\end{abstract}

\maketitle

\section{Introduction} \label{sec:introduction}
Artificial intelligence (AI) has become ubiquitous. Often manifested in machine learning algorithms, AI systems promise to be evermore present in everyday high-stakes tasks \citep{russell_artificial_2010, mitchell_machine_1997}. This is why building fair, responsible, and ethical systems is crucial to the design process of AI algorithms. Central to the topic of \textit{trusting} AI-generated results is the notion of \textit{interpretability}, also known as \textit{explainability}. The interpretability of AI models is a pivotal concern in contemporary AI research, particularly with the ubiquity of deep neural networks. This has given rise to research topics under the umbrella of interpretable machine learning (or IML) and explainable AI (or XAI), noting that the terms \textit{interpretable} and \textit{explainable} are used synonymously throughout the corresponding literature. Generically, interpretability is understood as the extent to which humans comprehend the output of an AI model that leads to decision-making \citep{molnar_interpretable_2022}. Humans strive to understand the ``thought process'' behind the decisions of the AI model --- otherwise, the system is referred to as a ``black box''. Even though the terms ``interpretation'' and ``explanation'' are used colloquially in the literature with varying degrees, throughout this paper, we use them as follows. \textit{Interpretable} entails a model that humans can understand and comprehend through direct observation of its internal workings or outputs. It implies that the model is intuitive to the human observer and that no further tools are required. On the other hand, \textit{explanation} refers to the output of a tool used to articulate the behavior of a model. In our case, and indeed many others, an explanation is a model itself, which is justifiably called such when it is interpretable. Complex models and their simpler explanations are often called black-box and white-box models, respectively.

\begin{figure}[ht]
    \centering    \includegraphics[width=0.99\columnwidth]{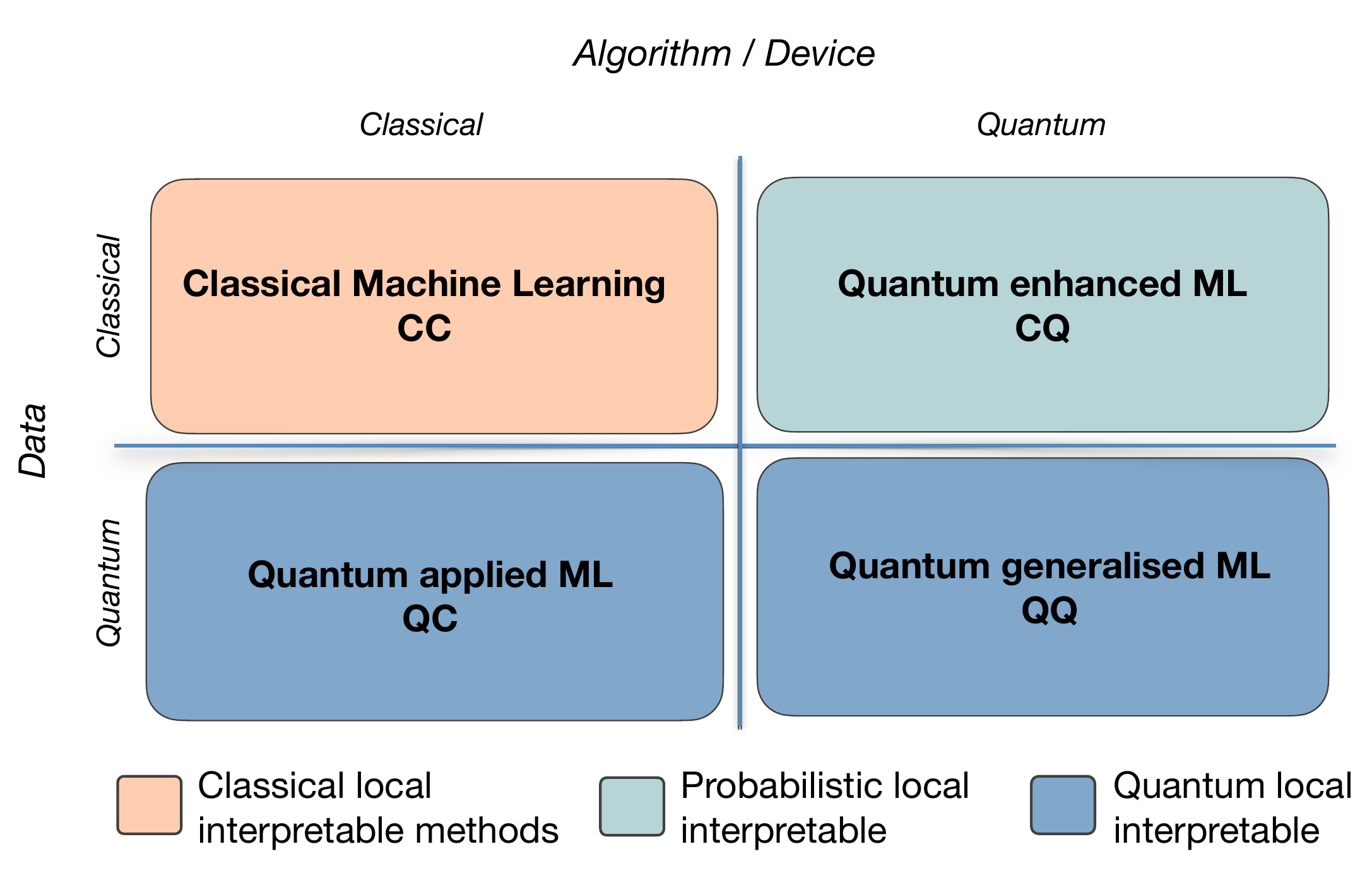}
    \caption{\textbf{Categorization of interpretability techniques as they apply to classical and quantum resources.} Here, the well-known QML diagram represents data and an algorithm or device, which can be classical (C) or quantum (Q) in four different scenarios. We consider a reformulation of interpretable techniques to be required in the CQ scenario. In the QC and QQ quadrants, the design of explicitly quantum interpretable methods may be required. The scope of this paper covers CQ approaches.}
    \label{fig:fig1}
\end{figure}

The precise definition of a model's interpretability has been the subject of much debate \citep{molnar_interpretable_2020, lipton_mythos_2018}. Naturally, there exist learning models which are more interpretable than others, such as simple decision trees. On the other hand, the models we prefer best for solving complex tasks, such as deep neural networks (DNNs), happen to be highly non-interpretable, which is due to their inherent non-linear layered architecture \citep{goodfellow_deep_2016}. We note that DNNs are one of the most widely used techniques in machine learning. Thus, the interpretability of neural networks is an essential topic within the field of interpretable ML research \cite{zhang_survey_2021, guidotti_survey_2018}. In this work, we focus on precisely the topic of interpretability as we consider the quantum side of neural networks.

In parallel, recent years have witnessed a surge of research efforts in \textit{quantum} machine learning (QML) \citep{biamonte_quantum_2017, schuld_machine_2021}. This research area sits at the intersection of machine learning and quantum computing. The development of QML has undergone different stages. Initially, the field started with the quest for speedups or quantum advantages. More recently, the target has morphed into further pursuits in expressivity and generalization power of quantum models. Nowadays, rather than ``competing'' with classical models, quantum models are further being enhanced on their own, which could, in turn, improve classical machine learning techniques. One of the key techniques currently used in QML research is the variational quantum algorithm, which acts as the quantum equivalent to classical neural networks \citep{cerezo_challenges_2022}. To clarify the analogy, we will refer to such models as quantum neural networks (QNNs) \citep{cerezo_variational_2021}. 

Given the close conceptual correspondence to classical neural networks, it is natural to analyze their interpretability, which is important for several reasons. Firstly, QNNs may complement classical AI algorithm design, making their interpretability at least as important as classical DNNs. Secondly, the quantum paradigms embedded into QNNs deserve to be understood and explained in their own right. The unique non-intuitive characteristics of quantum nature can make QNNs more complicated to interpret from the point of view of human understandability. Finally, with the growing interest and capabilities of quantum technologies, it is crucial to identify and mitigate potential sources of errors that plague conventional AI due to a lack of transparency.

In this work, we define and establish key notions pertaining to the interpretability of quantum neural networks. Through this formalization, we aim to contribute a robust foundation for advancing the discourse on the interpretability of quantum neural networks within the broader context of quantum machine learning research. In doing so, we generalize some well-known interpretable techniques to the quantum domain. Consider the standard relationship diagram in QML between data and algorithm (or device) type, where either can be classical (C) or quantum (Q). This entails the following combinations (CC, CQ, QC, and QQ), shown in Fig. \ref{fig:fig1}. Classical interpretable techniques are the apparent domain of CC. We will discuss, but not dwell on, the potential need for entirely new techniques when the data is quantum (QC and QQ). In CQ, the domain that covers the so-called quantum-enhanced machine learning techniques, although the data is classical, the output of the quantum devices is irreversibly probabilistic. CQ can be seen as a specialization of the quantum data techniques above. Generalizing classical notions of interpretability to this domain is the subject of our work. 

The question of interpretability in quantum machine learning models more broadly, as well as of QNNs more specifically, has already started to receive attention \citep{heese_explaining_2023, mercaldo_towards_2022, steinmuller_explainable_2022}, particularly involving the concept of Shapley values  \citep{lundberg_unified_2017}, which attempt to quantify the importance of features in making predictions. In \citep{heese_explaining_2023}, interpretability is explored using Shapley values for quantum models by quantifying the importance of each gate to a given prediction. The complexity of computing Shapley values for generalized quantum scenarios is analyzed in \citep{burge_quantum_2023}. In \citep{steinmuller_explainable_2022}, Shapley values are computed for a particular class of QNNs. Our work complements these efforts using an alternative notion of explainability to be discussed in detail next.

\section{Interpretability in AI}
\subsection{Taxonomy}\label{sub:taxonomy}
There are several layers to the design of interpretability techniques. To start, they can be \textit{model-specific} or \textit{model-agnostic}. As the name suggests, model-specific methods are more restrictive in terms of which models they can be used to explain. As such, they are designed to explain one single type of model. In contrast, model-agnostic methods allow for more flexibility in usage as they can be used on a wide range of model types. At large, model-agnostic methods can have a \textit{global} or \textit{local} interpretability dimension. Locality determines the scope of explanations with respect to the model. Interpretability at a global level explains the average predictions of the model as a whole. At the same time, local interpretability gives explanations at the level of each sample. In another axis, these techniques can be \textit{active} (inherently interpretable) or \textit{passive} (post-hoc). The state of these interpretable paradigms implies the level of involvement of interpretable techniques in the outcome of the other parameters. Active techniques change the structure of the model itself by leaning towards making it more interpretable. In contrast, passive methods explain the model outcome once the training has finished. In comparison to model-agnostic methods, which work with samples at large, there also exist example-based explanations that explain selected data samples from a dataset. An example of this method is the $k$-nearest neighbors models, which average the outcome of $k$ nearest selected points.

Other than the idea of building interpretable techniques, or more precisely, techniques that interpret various models, there exist models that are inherently interpretable. Such models include linear regression, logistic regression, naive Bayes classifiers, decision trees, and more. This feature makes them good candidates as surrogate models for interpretability. Based on this paradigm, the concept of surrogate models exists, which uses interpretable techniques as a building block for designing other interpretable methods. Such important techniques are, for example, local interpretable model-agnostic explanations (LIME) \citep{ribeiro_why_2016} and Shapley additive explanations (known as SHAP) \citep{lundberg_unified_2017}.

\subsection{Interpretability of neural networks}
The interpretability of neural networks remains a challenge on its own. This tends to amplify in complex models with many layers and parameters. Nevertheless, there is active research in the field and several proposed interpretable techniques \citep{zhang_survey_2021}. Such techniques that aim to gain insights into the decision-making of a neural network include saliency maps \citep{simonyan_deep_2013}, feature visualization \citep{olah_feature_2017, yosinski_understanding_2015}, perturbation or occlusion-based methods \citep{ribeiro_why_2016, lundberg_unified_2017}, and layerwise relevance propagation (also known by its acronym LRP) \citep{bach_pixel_2015}.

To expand further on the above-mentioned techniques, saliency maps use backpropagation and gradient information to identify the most influential regions contributing to the output result. This technique is also called pixel attribution \citep{molnar_interpretable_2020}. Feature visualization, particularly useful for convolutional natural networks, is a technique that analyses the importance of particular features in a dataset by visualizing the patterns that activate the output. In the same remark, in terms of network visualizations, Ref. \citep{zeiler_visualizing_2014} goes deeper into the layers of a convolutional neural network to gain an understanding of the features. This result, in particular, shows the intricacies and the rather intuitive process involved in the decision-making procedure of a network as it goes through deeper layers. Occlusion-based methods aim to perturb or manipulate certain parts of the data samples to observe how the explanations change. These methods are important in highlighting deeper issues in neural networks. Similarly, layerwise relevance propagation techniques re-assign importance weight to the input data by analyzing the output. This helps the understanding by providing a hypothesis over the output decision. Finally, the class of surrogate-based methods mentioned above is certainly applicable in neural networks as well.

The importance of these techniques is also beyond the interpretability measures for human understanding. They can also be seen as methods of debugging and thus improving the result of a neural network as in Ref. \citep{zeiler_visualizing_2014}. Below, we take a closer look at surrogate model-agnostic local interpretable techniques, which are applicable to DNNs as well.

\subsection{Local interpretable methods}
Local interpretable methods tend to focus on individual data samples of interest. One of these methods relies on explaining a black-box model using inherently interpretable models, also known as surrogate methods. These methods act as a bridge between the two model types. The prototype of these techniques is the so-called local interpretable model-agnostic explanations (LIME), which has received much attention since its invention in 2016 \citep{ribeiro_why_2016}. Local surrogate methods work by training an interpretable surrogate model that approximates the result of the black-box model to be explained. LIME, for instance, is categorized as a perturbation-based technique that perturbs the input dataset. Locality in LIME refers to the fact that the surrogate model is trained on the data point of interest, as opposed to the whole dataset (which would be the idea behind \textit{global} surrogate methods). Eq.~\eqref{eq:lime} represents the explanation $\xi$ of a sample $x$ via its two main terms, namely the term $L(f,g,\pi_x)$ representing the loss, which is the variable to be minimized, and $\Omega(g)$ which is the complexity measure, which encodes the degree of interpretability. Here $f$ is the black-box model, $g$ is the surrogate model, and $\pi_x$ defines the region in data space local to $x$ (See Alg. \ref{alg:lime} pseudocode). In broader terms, LIME is a trade-off between interpretability and accuracy,
\begin{equation}
    \xi (x) = \argmin_{g \in G} L(f,g,\pi_x) + \Omega(g) \label{eq:lime}
\end{equation}

In the following, we make use of the concept of local surrogacy to understand the interpretability of quantum models using LIME as a starting point. Much like LIME, we develop a \textit{framework} to provide explanations of black-box models in the quantum domain. The class of surrogate models, the locality measure, and the complexity measure are free parameters that must be specified and justified in each application of the framework.

\begin{algorithm}
\caption{LIME Algorithm \cite{ribeiro_why_2016}}\label{alg:lime}
\begin{algorithmic}[1]
\Function{LIME}{$f, x, D, K$}
    \For{$i = 1$ to $K$}
        \State Sample $z_i$ from $D$
        \State $g_i = f(z_i)$ \Comment{$g_i$ is the prediction of the classifier on the perturbed sample}
    \EndFor
    \State $\xi(x) = \argmin_{g \in G} L(f, g, \pi_x)$ \Comment{$L$ is a loss function, $\pi_x$ is a locality measure to the data point, and $G$ is the set of surrogate models}
    \State \textbf{return} $\xi(x)$
\EndFunction

\Function{Loss}{$f, g, \pi_x$}
    \State $L(f, g, \pi_x) = \sum_{z_i \in D} \pi_x(z_i) [f(z_i) \neq g(z_i)] + \Omega(g)$ \Comment{Penalize the difference between predictions}
    \State \textbf{return} $L(f, g, \pi_x)$
\EndFunction

\Function{Locality}{$\pi_x, z, x$}
    \State $\pi_x(z) = \exp\left(-\frac{d(x, z)^2}{\sigma^2}\right)$ \Comment{$d$ is a distance metric, $\sigma$ is a bandwidth parameter}
    \State \textbf{return} $\pi_x(z)$
\EndFunction
\end{algorithmic}
\end{algorithm}

\section{The case for quantum AI interpretability}\label{sec:qint}
As mentioned in Section \ref{sec:introduction}, interpretability in the quantum literature in the context of machine learning can take different directions. We consider the case when data is classical and encoded into a quantum state, which is manipulated by a variational quantum circuit before outputting a classical decision via quantum measurement. Our focus is on interpreting the classical output of the quantum model.

A quantum machine learning model $f$ takes as input data $x$ and first produces quantum data $|\psi(x)\rangle$. A trained quantum algorithm, such as the QNN, subsequently processes the quantum data and generates a classification decision based on the outcome of a quantum measurement. This is not conceptually different from a classical neural network beyond the fact that the weights and biases have been replaced by parameters of quantum processes, except for one crucial difference --- quantum measurements are unavoidably probabilistic.  

Probabilities, or quantities interpreted as such, often arise in conventional neural networks. However, these numbers are encoded in bits and directly accessible, so they are typically used to generate deterministic results (through thresholding, for example). Qubits, on the other hand, are not directly accessible. While procedures exist to reconstruct qubits from repeated measurements (generally called \textit{tomography}), these are inefficient --- defeating any purpose of encoding information into qubits in the first place. Hence, QML uniquely forces us to deal with uncertainty in interpreting its decisions.

In the case of probabilistic decisions, the notion of a \textit{decision boundary} is undefined. A reasonable alternative might be to define the boundary as those locations in data space where the classification is purely random (probability $\frac12$). A data point here is randomly assigned a label. For such a point, any \textit{explanation} for its label in a particular realization of the random decision process would be arbitrary and prone to error. It would be more accurate to admit that the algorithm is \textit{indecisive} at such a point. This rationale is equally valid for data points near such locations. Thus, we define the \textit{region of indecision} as follows,
\begin{equation}
    R = \left\{x' \in X: \left \lvert P(f(x')=1) - \frac{1}{2} \right \rvert < \epsilon\right\}, \label{eq:indecision}
\end{equation}
where $\epsilon$ is a small positive constant representing a threshold of uncertainty tolerated in the classification decision. In a certain context, points situated within this rationale lack a discernible \textit{rationale}, hence explanation, for their specific outcomes, relying instead on random circumstances.

At present, although certain data points possess labels that were randomly assigned, one may pose the question as to the underlying rationale for such assignments. In other words, even data points lying within the region of indecision demand an explanation. Next, we will show how the ideas of local interpretability can be extended to apply to the probabilistic or quantum settings. In doing so, we aim to provide a nuanced and comprehensive understanding of model predictions.

\subsection{Quantum LIME: Probabilistic local interpretability}

Here, we define Quantum LIME (or Q-LIME) as an extension of the classical counterpart to quantum models, namely quantum local interpretable model-agnostic explanations. A specialization of the definition of Q-LIME is the hybrid version in the $CQ$, such as in Fig. \ref{fig:fig1}.

In the context of the LIME algorithms, the loss function is typically chosen to compare models and their potential surrogates on a per-sample basis. However, if the model's output is random, the loss function will also be a random variable. An obvious strategy would be to define loss via expectation:
\begin{equation}
    \xi (x) = \argmin_{g \in G} \mathbb E [L(f,g,\pi_x)] + \Omega(g). \label{eq:lime_pr}
\end{equation}

Nevertheless, even under these circumstances,  a definitive assertion regarding $\xi$ as an explanation remains elusive. This stems from the recognition that its predictions are only capturing the average behavior of the underlying model's randomness. In fact, the label provided by $\xi$ may be the opposite of that assigned to $x$ by the model in any particular instance, which clearly does not capture a reasonable notion of \textit{explanation}.

To mitigate this, we generalize the classical definition and call an \textit{explanation} the distribution $\Xi$ of trained surrogate models $g$ obtained through the application of LIME to the random output of any probabilistic classifier. Note again that $g$ is random, trained on synthetic local data with random labels assigned by the underlying model. Thus, the explanation inherits any randomness from the underlying model. It is not the case that the explanation provides an interpretation of the randomness \textit{per se} --- however, we can utilize the distribution of surrogate models to simplify the region of indecision, hence providing an interpretation of it. 

The symbol $\Xi$ represents the Q-LIME explanation, emphasizing the interpretability framework tailored for QNNs. Additionally, Alg. \ref{alg:qlime} refers to a Monte Carlo approximation method employed within the Q-LIME process. This algorithm plays a pivotal role in facilitating the computation of interpretable insights by utilizing a Monte Carlo sampling approach, thereby enhancing the efficiency of the interpretation process.

\begin{algorithm}
\caption{Quantum LIME (Q-LIME)}\label{alg:qlime}
\begin{algorithmic}[1]
\Function{QLIME}{$U, x, D, K$}
    \State $\Xi \gets$ Empty list to store quantum surrogate models
    \For{$i = 1$ to $K$}
        \State $D_i \gets$ Generate synthetic quantum data locally around $x$
        \State $\gamma_i \gets$ \Call{LIME}{$U, x, D_i, K$} \Comment{Apply LIME to generate a quantum surrogate model (see Alg. 1)}
        \State Add $\gamma_i$ to $\Xi$
    \EndFor
    \State \textbf{return} $\Xi$
\EndFunction
\end{algorithmic}
\end{algorithm}

\subsection{Local region of indecision}

\begin{figure}
    \centering
    \includegraphics[width=0.40\textwidth]{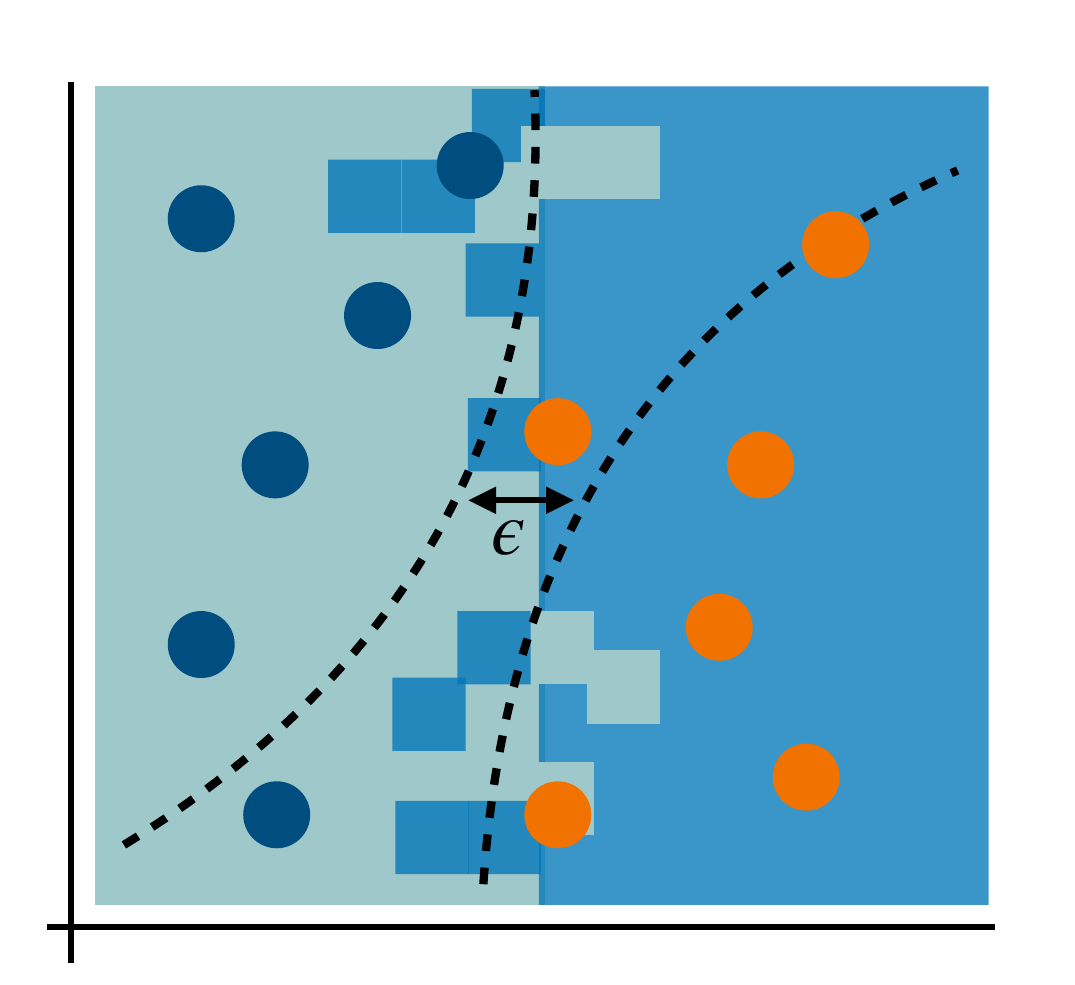}
    \caption{\textbf{Depiction of the concept of the \textit{local region of indecision}}. The space within the dashed lines represents the region in the decision space where data samples exhibit ambiguous classification due to randomness. The figure showcases a two-class classification task in a two-dimensional space with two features represented along the horizontal and vertical axes. Here, $\epsilon$ is the pre-defined threshold. Likely, data samples, either inside or close to the band, are interpreted as randomly assigned.}
    \label{fig:fig2}
\end{figure}

In this section, we define the \textit{local region of indecision}. In a broad sense, this is the region of indecision interpreted locally through a distribution of surrogate models. Suppose a particular data point lies within its own local region of indecision. The explanation for its label is thus the data was given a random label, even though its explanation may not be a satisfying one. Moreover, this is a strong statement because --- in principle --- all possible interpretable surrogate models have been considered in the optimization.

The local region of indecision can be defined as the region of the input space where the classification decision of the quantum model is uncertain. More formally, we can define the local region of indecision $B$ for a data point $x$ in a dataset $X$ as,

\begin{equation}
    B = \left\{x' \in X: \left \lvert P(g(x')=1|f,\Xi,\pi_x) - \frac{1}{2} \right \rvert < \epsilon\right\}, \label{eq:inexp}
\end{equation}
where $\epsilon$ is again a small positive constant representing a threshold of uncertainty tolerated in the classification decision --- this time with reference to the \textit{explanation} rather than the underlying model. 

The exact relationship between $B$ and $R$ is complicated, depending on the details of the loss function. However, for a $0-1$ loss function and $\pi_x$ defining a closed boundary around $x$, $B = R$ within $\pi_x$. In a more general case, $B$ approximates $R$ up to the freedom that elements of $\Xi$ are allowed to deviate from $f$. Note that the distribution in Eq. \eqref{eq:inexp} is over $\Xi$ as each $g$ provides deterministic labels. This has been formalized in Alg. \ref{alg:range}.

In much the same way that an interpretable model approximates decision boundaries locally in the classical context, a local region of indecision approximates the proper region of indecision in the quantum (or probabilistic) context. The size and shape of this region will depend on several factors, such as the choice of the interpretability technique, the complexity of the surrogate model, and the number of features in the dataset. We call the region a ``band'' as it describes the simplest schematic presented in Fig. \ref{fig:fig2}.

In the examples presented in this paper, we chose the interquartile range (IQR) as a summary measure for the set of quantum surrogate models $\Xi$. However, other options are undoubtedly possible. It is important to note that in higher-dimensional spaces, the complexity of quantum models and the diversity of decision boundaries may necessitate alternative summary measures.

\begin{algorithm}
\caption{Range}\label{alg:range}
\begin{algorithmic}[1]
\State $B \gets$ Empty list to store quantum decision boundaries
\State $\Xi \gets$ \Call{QLIME}{$U, x, D, K$} \Comment{Generate quantum surrogate models using Quantum LIME (see Alg. 2)}
\For{$i = 1$ to $M$} \Comment{Repeat for $M$ iterations}
    \State $\xi \gets$ Sample from $\Xi$
    \State $B_i \gets$ \Call{Boundary}{$\xi$} \Comment{Obtain the quantum decision boundary from the sampled quantum surrogate model}
    \State Add $B_i$ to $B$
\EndFor
\State $IQR \gets$ \Call{InterquartileRange}{$B$} \Comment{Compute the interquartile range of quantum decision boundaries}
\State \textbf{return} $IQR$
\Function{InterquartileRange}{$B$}
    \State Sort the quantum decision boundaries in $B$
    \State $Q_1 \gets$ The median of the lower half of $B$
    \State $Q_3 \gets$ The median of the upper half of $B$
    \State $IQR \gets Q_3 - Q_1$ \Comment{Interquartile range}
    \State \textbf{return} $IQR$
\EndFunction
\end{algorithmic}
\end{algorithm}

\section{Numerical experiments for interpreting QNNs}

We use the well-known Iris dataset \citep{fisher_use_1936} for our numerical experiments. To enhance the explicability of our methodology, we opt to simplify the analysis by framing it as a binary classification problem. This entails the exclusive utilization of two out of the three available classes within the dataset, coupled with the restriction to employ only two out of the four available features. In consideration of our indifference toward classifying the two categories of flowers, we abstract the names of these classes and labels below.

The trained quantum model to be explained is a hybrid QNN trained using simultaneous perturbation stochastic approximation (better known by its acronym SPSA) \citep{spall_overview_1998}, which is an optimization algorithm that efficiently estimates the gradient of an objective function. This model has been built and simulated using the Qiskit framework \cite{qiskit}. Each data point is encoded into a quantum state with the angle encoding  \citep{schuld_supervised_2018}, which acts by mapping numerical values to specific angles, facilitating their integration into quantum models. The QNN model is an autoencoder with alternating layers of single qubit rotations and entangling gates \cite{farhi_classification_2018}. Since our goal here is to illustrate the local region of indecision, as in Eq. \ref{eq:inexp}, we do not optimize over the complexity of surrogate models and instead fix our search to the class of logistic regression models with two features.

\begin{figure}[t!]
\includegraphics[width=0.99\columnwidth]{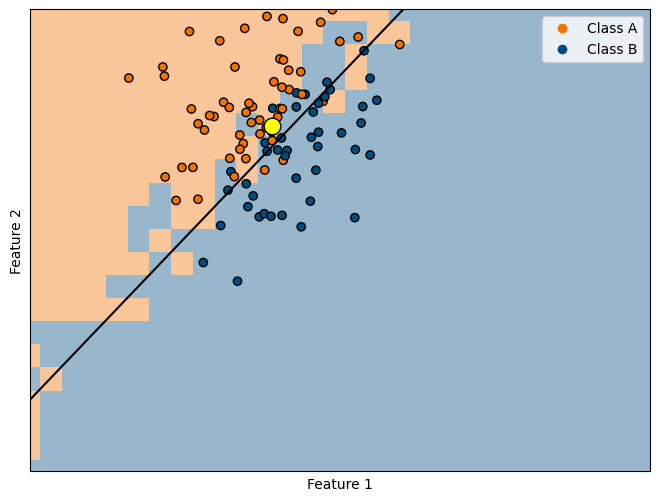}\\
\includegraphics[width=0.99\columnwidth]{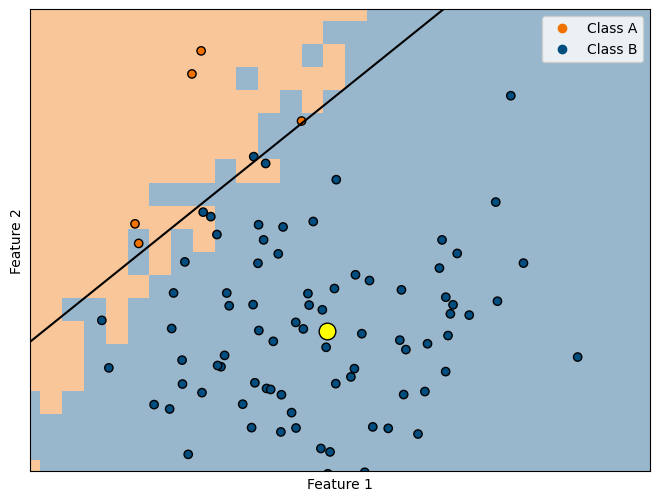}
\caption{\textbf{The classification uncertainty for a chosen data sample (yellow)}. The two shadings represent the decision boundary of the QNN, which is clearly randomly defined. The synthetic data and the produced surrogate linear decision boundary are shown for a single instance of QNN labels. (top) A clear example of an undefined classification result for the selected data point. (bottom) Decision boundary unambiguously separates the who classes with respect to the selected data point.}\label{fig:fig3}
\end{figure}

\begin{figure}[t!]
\includegraphics[width=0.99\columnwidth]{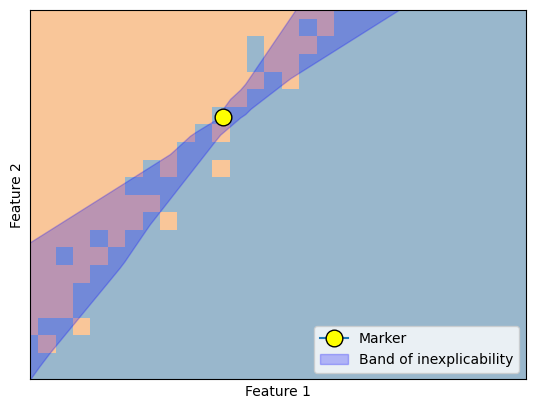}\\
\includegraphics[width=0.99\columnwidth]{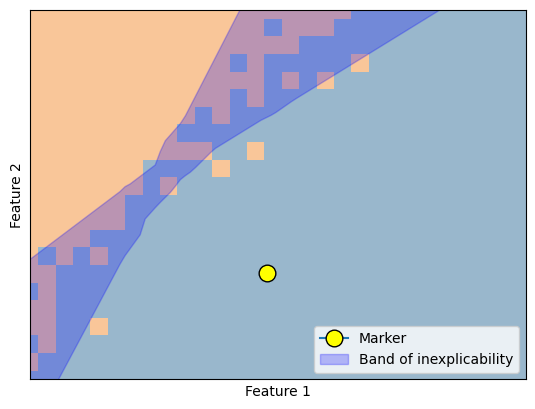}
\caption{\textbf{The approximated local region of indecision.} (top) An example of a marked data point that lies on the local region of indecision. (bottom) A data point that is outside of this region can be assessed for interpretability as per the interpretable techniques.}\label{fig:fig4}
\end{figure}

The shaded background in each plot of Figs. \ref{fig:fig3} and \ref{fig:fig4} show the decision region of the trained QNN. Upon inspection, it is clear that these decision regions, and the implied boundary, change with each execution of the QNN. In other words, the decision boundary is ill-defined. In Fig. \ref{fig:fig3}, we naively apply the LIME methodology to two data points --- one in the ambiguous region and one deep within the region corresponding to one of the labels. In the latter case, the output of the QNN is nearly deterministic in the local neighborhood of the chosen data point, and LIME works as expected --- it produces a consistent explanation. 

Nonetheless, in the first example, the data point receives a random label. It is clearly within the \textit{region of indecision} for a reasonably small choice of $\epsilon$. The ``explanation'' provided by LIME (summarized by its decision boundary shown as the solid line in Fig. \ref{fig:fig3}) is random. In other words, each application of LIME will produce a different explanation. For the chosen data point, the explanation itself produces opposite interpretations roughly half the time, and the predictions it makes are counter to the actual label provided by the QNN model to be explained roughly half the time. Clearly, this is an inappropriate situation to be applying such interpretability techniques. Heuristically, if a data point lies near the decision boundary of the surrogate model for QNN, we should not expect that it provides a satisfactory explanation for its label. Q-LIME and the \textit{local region of indecision} rectify this. 

For the same sample data points, the local region of indecision is shown in Fig. \ref{fig:fig4}. Data points within their own band should be regarded as having been classified due purely to chance --- and hence have no ``explanation'' in the conventional sense for the label they have been assigned. Note that the band itself is unlikely to yield an analytic form. Hence, some numerical approximation is required to calculate it in practice. Our approach, conceptually, was to repeat what was done to produce Fig. \ref{fig:fig3} many times and summarize the statistics to produce Fig. \ref{fig:fig4}. A more detailed description follows, and the implementation to reproduce the results presented here can be found at \cite{pira_local_2023}.

Approximating the local region of indecision requires a trained QNN, a sample data point, a function defining ``local'', and some meta-parameters defining the quality of the numerical approximation, such as the amount of synthetic data to be used and the number of Monte Carlo samples to take. Once these are defined, the general procedure is as follows, which is formalized in Alg. \ref{alg:qlime} and Alg. \ref{alg:range}.

\begin{enumerate}
    \item Create synthetic data locally around the chosen data point;
    \item Apply LIME to generate a surrogate model;
    \item Repeat 1 and 2 several times to produce an ensemble of decision boundaries.
    \item Take the interquartile range (for example) of those boundaries.
\end{enumerate}

To make evident, Fig. \ref{fig:fig3} is produced through a single application of steps 1 and 2 while Fig. \ref{fig:fig4} is produced after the final step. It is important to recognize that, in a practical or real-world application, it would be infeasible to plot the entire decision regions to any reasonable level of accuracy. The computational demands and complexities involved in plotting such extensive regions can be prohibitive. This makes the region of indecision both valuable and convenient in interpreting QNNs.

\section{Discussion \& Future Work}
In summary, in interpreting the classification output of a QNN, we have defined Q-LIME and the local region of indecision, a simple description of a region where the QNN is interpreted to be indecisive. The data samples within their associated band should not be expected to have an ``explanation'' in the deterministic classical sense. While directly useful for hybrid quantum-classical models, we hope this stimulates further research on the fundamental differences between quantum and classical model interpretability. In the remainder of the paper, we discuss possible future research directions.

Our results are pointed squarely at the randomness of quantum measurements, which might suggest that they are ``backward compatible'' with classical models. Indeed, randomness is present in training classical DNNs due to random weight initialization or the optimization techniques used. However, this type of randomness can be ``controlled'' via the concept of a \textit{seed}. Moreover, the penultimate output of DNNs (just before label assignment) is often a probability distribution. Nevertheless, these are produced through arbitrary choices of the activation function (i.e., the Softmax function), which force the output of a vector that merely \textit{resembles} a probability distribution. Each case of randomness in classical DNNs is very different from the innate and unavoidable randomness of quantum models. While our techniques \textit{could} be applied in a classical setting, the conclusions drawn from them may ironically be more complicated to justifiably action. 

In this work, we have provided concrete definitions for QNNs applied to binary classification problems. Using a probability of $\frac12$ would not be a suitable reference point in multi-class decision problems. There are many avenues to generalize our definitions, which would mirror standard generalizations in the move from binary to multi-class decision problems. One such example would be defining the region of indecision as that with nearly maximum entropy. 

We took as an act of brevity the omission of the word \textit{local} in many places where it may play a pivotal role. For example, the strongest conclusion our technique can reach is that a QNN is merely \textit{locally} inexplicable in the classical sense. In such cases, we could concede that (for some regions of data space) the behavior of QNN is inherently stochastic, lacking any discernible patterns or deterministic explanations. Alternatively, we can use this conclusion to signal that an explanation at a higher level of abstraction is required. This shift toward higher-level abstraction becomes imperative for gaining more insights into QNNs, acknowledging their inherent quantum nature and the limitations imposed by classical interpretability paradigms.

Classically, should a data point inquire, "Why has this label been assigned to me?", within a quantum framework, an initial response may invoke the inherent stochastic nature of quantum systems. Nonetheless, persistent interrogation from the data point could necessitate a quantum extension of \textit{global} interpretability methodologies to clarify the specific features or attributes contributing to the assigned label.

Referring back to Fig. \ref{fig:fig1}, we have focused here on CQ quantum machine learning models. Nevertheless, the core idea behind local surrogate models remains applicable in the context of quantum data --- use interpretable \textit{quantum} models as surrogate models to explain black box models producing quantum data. Of course, one of our assumptions is the parallels between the classical interpretable models we mentioned above, with their quantum equivalents. This can be a line for future work. The ideas here encapsulate inherently quantum models such as matrix product states or tensor network states, which can act as surrogate models for quantum models as they may be considered more interpretable. These surrogate models can serve as interpretable proxies, facilitating a comprehension of the underlying quantum processes, thus paving the way for enhanced interpretability of quantum models.

Furthermore, the idea behind \textit{interpreting} or ``opening up'' black-box models may be of interest in control theory \citep{youssry_experimental_2023, youssry_characterization_2020, youssry_noise_2023}. In this scenario, the concept of ``grey-box'' models --- portions of which encode specific physical models --- give insights into how to engineer certain parameters in a system. These grey-box models can thus be considered \textit{partially explainable} models. The proposed algorithm in \cite{weitz_sub_2023} may also be of interest in terms of creating intrinsically quantum interpretable models, which would act as surrogates for other more complex quantum models.

Advanced ideas for future work entail exploring and developing specific metrics tailored for assessing interpretability in QML models. This may involve defining measures that capture the degree of coherence or entanglement contributing to model predictions.

An obvious open question that inspires future research remains to investigate the difference in computational tractability of interpretability methods in quantum versus classical. This will lead to understanding whether it is more difficult to interpret quantum models as opposed to classical models. We hope such results shed light on more philosophical questions as well, such as \textit{is inexplicability, viz. complexity, necessary for learning?}. This exploration holds the potential to advance our understanding not only of the interpretability challenges in quantum models, but also of the fundamental relationship between complexity and the learning process itself.

For completeness, the case for the interpretability of machine learning models does not go without critique. Certain viewpoints contend that the pursuit of insights into a model's decision-making process should not come at the expense of sacrificing performance, and, in practical terms, this prioritization might not always be feasible or deemed necessary \citep{sarkar_explainable_2022, mccoy_believing_2022}. It is often posited that simpler models inherently possess a greater degree of explainability. Nonetheless, it is the more complex models that require explanations, as they may be more likely employed in critical applications.

Regardless of the two distinct camps of beliefs, the niche field of interpretable machine learning keeps growing in volume. An argument is that having a more complete picture of the model's performance can help improve the performance of the model overall. As QML becomes increasingly relevant to AI research, we expect the need for quantum interpretability will also be in demand. This research contributes to a broader understanding of explicability in quantum AI models, paving the way for the development of accountable and transparent systems in the quantum computing and AI era.

{\em Acknowledgments:} LP was supported by the Sydney Quantum Academy, Sydney, NSW, Australia.






\bibliography{interpret.bib}

\end{document}